\documentclass[runningheads]{cl2emult}

\usepackage{makeidx}  
\usepackage{graphicx} 
\usepackage{subeqnar} 
\usepackage{multicol} 
\usepackage{cropmark} 
\usepackage{eso}      
\makeindex            

\begin{document}
\title*{What May We Learn from Multi-wavelength \protect\newline Observations of Active Galactic Nuclei}
\toctitle{What May We Learn from Multi-wavelength \protect\newline Observations of Active Galactic Nuclei}
%
%
\titlerunning{Multi-wavelength  Observations of Active Galactic Nuclei}
%
\author{Thierry J.-L. Courvoisier\inst{1}}
%
\authorrunning{Thierry J.-L. Courvoisier}
%
%
\institute{INTEGRAL Science Data centre, 16 ch. d'Ecogia, CH-1290 VERSOIX, Switzerland \protect\newline and \protect\newline Observatoire de Gen\`eve, CH1290 SAUVERNY, Switzerland}
\maketitle              

\begin{abstract}
We discuss how several of the questions that remain unclear on the physics of Active Galactic Nuclei may find elements of answers when using in the coming years the extraordinary set of instruments that will be flying simultaneously to observe in all bands of the electromagnetic spectrum. The choice of questions mentioned here is personal and not exhaustive.

\end{abstract}

\section{What do we know about Active Galactic Nuclei?}
Wondering what may be learned about Active Galactic Nuclei (AGN) in the coming years requires that one thinks about what is currently known and about the type of knowledge that is relevant. Indeed, AGN seem to be complex systems that need  not be described at a level at which the general physical processes are hidden behind local and/or statistical fluctuations. The learning process may in some sense be similar to understanding  the physics of the terrestrial atmosphere by doing meteorological observations. The aim in this case is not to describe the  local weather but to extract from the complex descriptions of the observations the guiding principles. Clearly, what knowledge or understanding one wishes to obtain on any given subject is a matter of personal taste. Indeed many people, even among astronomers, might have only a limited interest in AGN. The following account is therefore a personal approach to the subject.

\vspace{1cm}
Among the facts that are well established now and documented in a number of books and reviews in recent years one finds:

\begin{description}
\item[The source of energy] 
is the gravitational energy of matter falling into a massive black hole. 

With this general principle one deduces two  characteristics of AGN, namely an estimate of the mass of the central black hole through the Eddington luminosity (more appropriately, this estimate gives a lower boundary to these masses). This mass  is found to be up to $10^{10}$ solar masses. Also from the luminosity of the objects and from reasonable assumptions on the efficiency with which accreted matter can be made available to be radiated one deduces the rate at which matter must be accreted into the central black hole. Rates of up to 100 solar masses per year for the brightest nuclei are found. 

The role of the rotation of the central black hole in the global energetics still needs be understood. 

\item[The emission processes] through which energy is radiated are now mostly identified. They include synchrotron emission, emission of dust heated by ultraviolet and X-ray radiation fields, Compton processes (be they thermal or due to relativistic electron distributions) and emission from optically thick media (black body radiation). It is  possible from the identification of the emission processes and the shape of the spectral energy distributions to deduce (some of) the physical conditions of the emitting material.
\item[The state of the gas] is known for the gas that emits the lines charateristics of most AGN types. Temperature, density and filling factors as well as the origin of the ionisation can be read from the spectra and provide a picture in which photoionisation plays a dominant  role.
\item[Aberrations and geometry] play  an important role through Doppler boosting of components related to relativistic jets and through absorption of some components that happen to lie behind dense clouds of dust and gas. These effects considerably modify the appearance of AGN and probably lead to our AGN classification scheme.
\item[AGN sit in the center of galaxies] whith which they must have intense relations, be it only through material that is falling into the galactic potential well to the AGN.
\item[The evolution] of AGN is well documented. The AGN  number density and/or luminosity have greatly changed over the age of the Universe.
\item[AGN have a very rich phenomenology] for which we have large amounts of data. We thus have descriptions of many spectral energy distributions, of many types of variations in different objects and at different wavelengths and we have a vast zoology of objects in which populations differ greatly in one or several characteristic parameters. We are not always succesful in understanding this rich phenomenology in terms of underlying physical processes.
\end{description}

\section{Current questions}

\subsection{Reprocessing}

We know that reprocessing of radiation by a medium takes place in AGN. Examples are the emission of heated dust (reprocessed UV radiation), Compton processes in which soft photons aquire additional energy from hot electrons and the presence of a fluorescence Fe line among others. 

What still needs be quantitatively assessed is the relative importance of the so-called Compton reflection hump that was suggested by \cite{Poundsetal90} to be important between 10 and some tens of KeV. This hump is due to the reflection of a primary X-ray component (taken as a power law) by cold material. That this causes a hump can be qualitatively understood by noting that the primary component is absorbed by the medium at low energies (the medium is assumed optically thick) while at high energies the scattering cross section decreases due to Klein-Nishina effects; see \cite{MDP93} for a review. There will therefore be a region in the spectrum where the reflection is maximal, thus producing a broad hump on top of the primary component.

It is suggested in models of the ultraviolet emission of AGN (see next section) that the UV emission is due to a disk heated by the primary X-ray component rather than by gravitational energy released within the (optically thick) accretion disk. In this scenario the reflection hump observed in the medium energy X-rays, the fluorescence line and the UV emission are all due to the same reprocessing scheme. All three elements should be consistently observed and variations in these components must be closely correlated and related (albeit possibly in a complex way) to the variations of the primary X-ray component.

Observations spanning the ultraviolet to hard X-ray domain can provide measurements that will allow us to separate the primary component from the reprocessed ones, to measure the flux in all components as a function of time and to seek some geometrical information by looking at the relative intensity of the components in different classes and as a function of time.

These observations will be possible using XMM and INTEGRAL in a coordinated manner over the coming decade on a variety of objects.

\subsection{The Blue Bump}

The peak in emission in the blue-ultraviolet region of the spectrum that is observed in Seyfert galaxies and quasars has been associated with the accretion process since 1978 \cite{Shields78}. This suggestion has given rise to a large activity in the subsequent years to provide a good match beteen observations and accretion disk theory. Despite 2 decades of work the concordance between the observations and theory is still not satisfactory. The continuum spectral energy distribution, spectral features, polarisation, relationship between continuum shape and luminosity and, possibly most importantly, the variability properties of the blue bump do not match those expected from an optically thick medium inside which gravitational energy is released to be radiated at the surface of the medium; see \cite{CC91} and the review on  AGN accretion disk related problems in  \cite{KB99}.

The blue bump flux is observed to vary almost simultaneously at all wavelengths from the visible to the ultraviolet range in 3C~273 as well as in Seyfert galaxies \cite{CC91}. This is in contradiction with the expectations that disturbances that propagate through the accretion disk (or whatever other structure) do so on viscous timescales that are many order of magnitude longer than the observed lags or upper limits thereof. This has lead to the development of models in which an X-ray source shines onto the disk. The latter is thus heated from the outside and the illuminated regions respond to variations of the primary source on timescales that are given by light propagation effects in good agreement with measurements of lags \cite{PCW98}. One of the most elaborate models of this type is that of \cite{HMG94} in which a multiple corona shines onto
the disk. It may be worth mentioning that the observations require only that the signal responsible for the perturbances propagate with the speed of light, not necessarily that the energy be distributed with the speed of light. 

In a detailed study of the visible and ultraviolet emission of 3C~273 \cite{PCW98} have  shown that the short wavelength emission of the blue bump in 3C~273 can be well modeled by a series of independent events occuring at a rate of about 3 per year and having each an energy of some $10^{52}$\,ergs. If this approach is physically sound (a good mathematical description of the data is not sufficient to ascertain the physical validity of the assumptions) than several questions must be answered: What is the origin of the events (see  next subsection for a proposal)? What about the long wavelength emission of the blue bump? Can the analysis be generalised to other objects?

Be it in the frame of the accretion disk models with or without a corona or in that of the individual events, there are still fundamental questions to be solved to explain the nature of the blue bump emission.

\subsection{How Does the Accretion Proceed?}

Accretion disks \cite{SS73} are a natural gas accretion structure in that they provide a means of expelling angular momentum while the material is being accreted  provided that the nature of the viscosity is understood. 

The difficulties of the accretion disk models to explain the observed properties of the blue bump, the description of the blue bump in 3C~273 in terms of individual events and the presence of stars in the center of galaxies has lead us to study a model in which the accreted material is not a diffuse gas but rather in the form of stars \cite{CPW96}, \cite{T00}. Stars orbiting a supermassive black hole at some 100 gravitational radii have velocities close to 0.1\,c. When two such stars collide their kinetic energy amounts to some $10^{52}$\,ergs for stars of the mass of the Sun. This is in good agreement with the energy contained in one event as discussed in the previous section.

In order to have few collisions per year (to provide the event rate described above) and to explain the average luminosity of the quasars there must be approximately the same mass in the central black hole and in the stellar population within some 100's of gravitational radii. In this volume the star density must be therefore very high, of the order of $10^{11}-10^{14}$ stars per cubic parsec. 

Accretion of matter in the form of stars and dissipation of the kinetic energy through stellar collisions rather than in a disk  has the property that the dependence of the variability on the object luminosity  discussed in \cite{PC97} based on IUE observations can be explained. This dependence is much flatter than expected based on independent events that are Poissonian distributed in time, suggesting that the event properties are a function of the average luminosity of the object. The star accretion scenario provides this naturally, because the average star collision
energies depend on the distribution of stars with distance to the black hole, which also determines the average luminosity of the object \cite{T00}.

Much work is still needed to assess whether this accretion process has a good prospect to contribute to the luminosity of AGN in a significant way. The emission processes occuring during and after a collision must be calculated and compared with observations of spectral energy densities and variability, the dynamics and evolution of very  dense clusters in the vicinity of a massive black hole must be studied. An alternative line of work may be to  prove that this model is irrelevant by showing that the accretion material is organised in a single plane and rotating all in the same direction.

\subsection{Links Between Starburst and AGN Activities?}

Starburst activity is observed in relation with AGN (the luminosity of which comes from accretion processes). This type of link can intuitively be expected. As the material falls in the central regions of a galaxy it will dissipate angular momentum. This will stir the interstellar material and provide conditions that are propitious for star forming activity.

Starburst activity with a luminosity that is comparable to that of the AGN seems to be common among absorbed Seyfert galaxies (Seyfert II galaxies). Both \cite{O99} and \cite{DH99} report that at least close to half of the Seyfert II galaxies that they observed have important starburst activity. On the other hand, \cite{O99} report that none of the 8 unobscured Seyfert galaxies (Seyfert I galaxies) they have observed  showed starburst properties.

That the difference between Seyfert I and Seyfert II galaxies lies in the presence of absorbing material that obscures the nuclei of the Seyfert II galaxies is in no doubt based on the observation of broad lines in the polarised light of Seyfert II galaxies \cite{AM85} and the clearly absorbed
X-ray emission of the same. 

The observations described above, suggest further that the presence of the obscuring material is related to the star forming activity. This link may indeed be expected as star formation requires the presence of large quantities of gas and dust forming dense clouds that are also associated with higher probabilities of obscuration.

The observations discussed here are in marked contrast with the expectations of the unified models \cite{A93} which postulate that Seyfert I and Seyfert II galaxies are differentiated by their orientation with respect to the line of sight. The model suggested by the different level of starburst activity in both types of galaxies is indeed that the nuclei are similar but that the presence of obscuring material is a genuine difference in the two classes of objects. The role of orientation effects is then probably secondary and needs be studied with the help of complete samples.

\subsection{Complete Samples}

AGN and quasars come in a great variety of disguises that led to a complex classification. This is due in great part to the fact that some properties are enhanced when associated with a relativistic jet pointed towards us (and correspondingly weakened when the jet is pointed in other directions) while some other aspects may be hidden by obscuring material as noted in the previous subsection. It is worth noting that some of the differences are certainly intrinsic to the objects and not simply related to geometrical and/or Doppler boosting and/or absorption. Among those intrinsic differences is the presence of the blue bump and the existence of the jet in the first place.

In order to understand what are the intrinsic differences as opposed to viewing or orientation effects and in order to measure the effects related to viewing and orientation it is mandatory to obtain complete samples of AGN. This task is proving difficult because of the difficulty to find parameters that are indeed independent of orientation and absorption and that are unique signatures of AGN activity. The presence of hard X-ray radiation is for example a clear signature of AGN activity and, above the energy at which absorption is important, is rather insensitive to the presence of obscuration material. The X-ray emission of radio loud quasars is, however, more important than that of radio quiet quasars \cite{P00}. This may be due to the fact that a fraction of the X-ray emission is linked to relativistic jets. Thus samples based on hard X-ray flux measurements of AGN whilst unbiased towards obscured or unobscured AGN will contain biases related to the presence or not of jets and to their orientation with respect to the line of sight. 

\cite{P00} have computed the continuum spectral energy distribution of a small sample of quasars with very different radio properties for which far infrared data from ISO have been measured. This analysis which is based on a sample that is not complete suggests that the far infrared emission is the emission component that is least affected by biases. This may indeed be expected since dust emission occurs on  large scales, is isotropic and is unrelated to the jet. Nonetheless if confirmed this implies that the dust properties of quasars are independent of their class, a non trivial result. Based on this preliminary result we suggest that, at least for quasars, complete samples should be based on far infrared measurements of objects that are identified by their X-ray emission to be AGN rather than starburst galaxies. This approach may not work in the case of BL Lac objects in which the far infrared emission is dominated by synchrotron emission.

\subsection{AGN Versus Stars as Photon Sources in the Universe}

Several authors (also in this volume) have noted the similar shapes of the star formation history in the Universe and of the QSO or AGN evolution. This similarity is then taken as a strong suggestion that star formation and AGN activity are intimately linked in the development of galaxies.

\cite{D98} has compared the the radio luminosity of quasars as a function of redshift with the star formation history in the Universe and has come to the conclusion that there are $10^7$ solar masses of stars formed per solar mass accreted into a black hole in a radio loud quasar. As far as energetics is concerned a 10 solar mass star is representative of a  population and generates about $4 \cdot 10^{52}$ ergs over its lifetime (Maeder private communication). This means that for $10^7$ solar masses some $4 \cdot 10^{58}$\, ergs are generated. The energy liberated by the accretion of one solar mass assuming an efficiency of 10\% is $2 \cdot 10^{53}$ ergs. It follows that the
ratio of accreted energy generation onto massive black holes in radio loud quasars to that of nucleosynthesis in the Universe is $10^{-5}$. In order to convert this ratio into a global estimate of the total rate of energy production by accretion onto all massive black holes to that due to nucleosynthesis it would be necessary to know the  density ratio of  AGN to the radio loud QSO's used in Dunlop (1998) and their respective average luminosities. 

\cite{FI99} have used the hard X-ray background and generic AGN spectral energy distributions to estimate the total contribution of AGN to the photon flux and compared this with the photon flux of star formation. They conclude that the total power emitted by accretion is about 1/5 that due to stars in the Universe. A further comparison between star formation history and quasar evolution is presented in \cite{Fetal99} and can also be used to derive that the ratio of accretion to star formation power is of the order of $10^{-1}$.

The close ties between the star formation and AGN luminosity in the Universe as it evolved provides a fascinating example of the interconnection of phenomena which many of us would have thought independent.

\subsection{What Makes an AGN Radio Loud?}

A small fraction of AGN are radio loud. It would indeed be interesting to know this fraction with some precision as a function of the luminosity and redshift of the AGN. The orgin of this difference cannot be sought in absorption as radio waves are undisturbed by neutral gas and dust, nor can it be sought in orientation and Doppler boosting effects, because  radio galaxies show that a significant fraction of the radio luminosity is emitted isotropically. The presence or absence (or at least extreme weakness) of radio emission is therefore intrinsic to the objects.

Radio emission is related to the presence of jets in the AGN. The origin of the radio emission is therefore related to the presence of jets. The question may therefore be re-phrased to asking why a small fraction of AGN have jets and others not. The presence of magnetic fields and/or the rotation of the black hole certainly play a role in these question.

\subsection{What Turns a Galaxy in an AGN?}

Evidence for black holes in galaxies has now become strong, in particular in the center of our Galaxy \cite{EG00}. The luminosity of the galactic center, and that of the nuclei of other galaxies in which black holes are probable, is, however, orders of magnitude less than the Eddington luminosities inferred from the masses of the black holes . The presence of the black hole  is, therefore, not a sufficient condition for the presence of a bright AGN in a galaxy. There must be in addition to the black hole  a sufficient quantity of matter that can be accreted in order to turn the central black hole of a galaxy into an active AGN. This matter can be in the form of diffuse gas accreted through a disk or in the form of a sufficiently dense cluster of stars surrounding the black hole.

\subsection{What holds the Broad Line Region Clouds together?}

\cite{Detal99} have recently shown by observing with high resolution and high signal to noise the broad lines of 3C~273 that the number of clouds is in excess of $10^8$. This number is larger than the number of stars that have been suggested as a source of confinement for the clouds \cite{SN88}, \cite{AN94}, \cite{AN97}. It is therefore probable that models in which the broad line region is seen as a continous hydrodynamic medium will have to be preferred to those in which  a collection of well defined clouds exist for a significant time. See \cite{Detal99} for a list of references to such models.

\section{Multi-Wavelength Observations}

All the questions discussed above have to be addressed with observations in several wavebands. This is even true for the question of the broad line cloud confinement in which a possible confining medium could have been detected through X-ray observations.

We will be in a very priviliged position in the next years with respect to multi-waveband coordinated observations of AGN with the simultaneous presence in orbit of high energy observing satellites that will cover the energy domain from 10s of GeV (AGILE, AMS) to MeV and 100s of keV (INTEGRAL), to soft X-rays and ultraviolet (XMM, Chandra, Spektrum X-G). These high energy instruments are complemented with a large array of very performing ground based telescopes in the visible, near infrared and radio domains of the spectrum and of the satellite  SIRTF in the far infrared domain. We should thus be in a position to address many of the questions above in a very comprehensive way.

One can reasonably expect that well planned observations using all these facilities should allow us to solve the question of re-processing, that of the relationship between starburst and AGN activity and to obtain well defined complete samples  in a convincing way. We should then be able to make very significant advances on many of the other topics mentionned here and in particular on the way in which material is accreted and the nature of the blue bump.

To be able to, and to have to, formulate the questions mentionned here is both a tribute to those who have pioneered the field of AGN, Lo Woltjer being prominent among them, and a sign that the subject is difficult and has been progressing only slowly once the bases were laid. It is our hope that the terms in which the questions are expressed and the tools that are now available will allow us to progress. We are well aware, though, that we are not the first to express these hopes and that observations of AGN have often provided new questions rather than answers to those they were expected to solve.

\clearpage
\addcontentsline{toc}{section}{Index}
\flushbottom
\printindex

\end{document}